\documentclass{elsart}
\usepackage{epsfig}

\begin{document}

\begin{frontmatter}

\title{\Large \bf Neutron beta decay and the current determination of 
$V_{ud}$.}

\author{A. Garc\'\i a},
\author{J.L. Garc\'\i a--Luna\thanksref{jlg}} and 
\author{G. L\'opez Castro \thanksref{glo}}
\address{Departamento  F\'\i sica, Centro de Investigaci\'on y de
Estudios Avanzados del IPN, Apdo. Postal 14-740, 07000 M\'exico,
D.F., M\'exico}
\thanks [jlg] {E--mail: jlgarcia@fis.cinvestav.mx}
\thanks [glo] {E--mail: glopez@fis.cinvestav.mx}

\maketitle 

\begin{abstract} 

Measurements of neutron beta decay observables required to determine
$|V_{ud}|$ are reaching the 0.1\% accuracy. In this paper we review the
calculation of the decay rate of this process, discuss its relevant
uncertainties, and obtain an expression
that is precise at the $10^{-4}$ level. Our analysis clearly shows the necessity of more precise
measurements of $\lambda$, the ratio of axial/vector couplings. The current
situation in neutron beta decay is that one cannot yet quote a single consistent
value for $|V_{ud}|$ from it. We also
discuss the region of parameter space in the $|V_{ud}|$--$\lambda$ plane
where new physics effects should lie, if they contribute to neutron beta
decay.
\end{abstract} 

\end{frontmatter}

{\large \bf 1. Introduction.} Neutron beta decay can provide in the future the most precise determination of the Cabibbo-Kobayashi-Maskawa
(CKM) matrix element $|V_{ud}|$ \cite{ckm}. The experimental inputs
required for this purpose are the neutron lifetime $\tau_n$ and  the
ratio of axial-vector to vector couplings, $\lambda \equiv g_1(0)/f_1(0)$. An 
update of the precise  computation of the neutron decay rate is
important for (at least) the following reasons: (a) the measurement of
$\tau_n$ and $\lambda$ have reached the 0.1\% accuracy and further
improvements are expected in the future (see for example \cite{deutsch})
and (b) independent and precise determination of $V_{ud}$ is compelling to 
test the unitarity condition $\sum_{i=d,s,b}|V_{ui}|^2=1$ of the CKM matrix. 
This requires from theory the
computation of the decay rate at the $10^{-4}$ level in accuracy, and will 
allow a determination of $|V_{ud}|$ with a precision $0.1\%$ which
becomes competitive with the determination coming from superallowed
Fermi transitions (SFT) among $J^P=0^+$ nuclei \cite{pdg98}, namely,
\begin{equation}
|V_{ud}| =0.9740 \pm 0.0010\ .
\label{vud}
\end{equation}
 The purpose of the present paper is to review the computation of the
neutron beta decay rate and discuss its relevant uncertainties at the $10^{-4}$ 
level. At this level it becomes necessary to discuss the relevance of the induced hadronic
form factors and even radiative corrections up to $\alpha^2$ 
contributions \cite{wilkinson1,gk}.\\ 
\hspace*{.5cm}There are two reasons why $|V_{ud}|$ can be
obtained from SFT with very small error bars. One is of statistical nature,
because Eq.(\ref{vud}) arises from the weighted average of 9 measured transitions 
\cite{th1}. The other one is an advantage of SFT over neutron
beta decay; they involve at tree-level only the vector current. The present
uncertainty in $|V_{ud}|$ from SFT has two dominant (theoretical) sources: the
nuclear-dependent (mismatch) isospin breaking corrections and the piece of
photonic corrections induced by the axial-vector current \cite{ms1,towner}. The
advantage of neutron beta decay is that it is free of a nuclear environment, but
its (theoretical) disadvantage is that it involves at tree-level the
axial-vector current, too. Away from a nuclear environment, the 
isospin-breaking corrections are more reliable to compute. Such corrections 
clearly affect both leading
form factors. However, the isospin breaking corrections enter into $f_1(0)$ only
at second order in the breaking by virtue of the Ademollo-Gatto theorem 
\cite{ademollo}. This fact makes them very small and they contribute only at the 
$10^{-5}$ level \cite{paver}. On the other hand, these corrections are more
important in $g_1(0)$ and, also, they are more difficult to compute reliably.
This complication is avoided using the experimental value of $g_1(0)$, through
the measurement of the ratio $\lambda$. The main source for such measurements is
the electron-neutron-spin asymmetry coefficient. It is because the experimental
uncertainty in $\lambda$ can be much improved over its theoretical uncertainty
that the neutron beta decay determination of $|V_{ud}|$ may become its
best determination.\\
\hspace*{.5cm}For sakeness of clarity we will present the different ingredients for
the determination of the decay rate in the following order. In section 2 we
provide the expression for the decay rate of the neutron at tree-level by
including small effects of non-leading form factors (other than $f_1(0)$
and $g_1(0)$). In section 3 we address the issue
of radiative corrections. We will discuss the uncertainties associated with
factorization of different orders in $\alpha$ as commonly presented in
the literature. In section 4 we summarize the current information of
experimental data on $\tau_n$ and $\lambda$ including the most recent
results. In particular, we will discuss the discrepancies in different 
measurements of $\lambda$, and we will provide a comparison of the different 
determinations of $|V_{ud}|$ as obtained from different sources. Section 5 
contains our conclusions.\\
{ \large \bf 2. Decay rate at the tree-level.} The tree-level amplitude
(meaning that radiative corrections are ignored) for the decay $n \rightarrow p e^- \bar\nu_e$
can be parametrized in terms of six Lorentz-invariant form factors. The
most general form of this amplitude is given by (we follow the conventions of 
Ref. \cite{gk}):
\begin{eqnarray}
 M^0  &=& \frac{G_0}{\sqrt{2}} V_{ud} \bar u(p')\left\{
f_1(q^2)\gamma_{\mu}+\frac{f_2(q^2)}{m_n}\sigma_{\mu\nu}q^{\nu} 
+\frac{f_3(q^2)}{m_n}q_{\mu}\right. \nonumber \\ 
&& \left. + \left[ g_1(q^2)\gamma_{\mu}
+\frac{g_2(q^2)}{m_n}\sigma_{\mu\nu}q^{\nu} +\frac{g_3(q^2)}{m_n}q_{\mu}
\right] \gamma_5   \right\} u(p) l^{\mu}\ ,
\label{ampli} 
\end{eqnarray}
where $G_0$ is the effective weak coupling of the relevant four-fermion 
hamiltonian at the tree-level  and $l_{\mu}$ is the leptonic V-A current. 
$f_i, g_i$ denote dimensionless $q^2=(p-p')^2$ dependent form factors. Since 
the momentum transfer is very small ($q^2_{max}/m_n^2\approx 1.9 
\times 10^{-6}$), we can already anticipate that the contributions 
of the form factors $f_2, f_3, g_2$ and $g_3$ to the decay rate are very 
suppressed. By the same reason, the $q^2$ dependence of form factors can be 
safely neglected.\\
\hspace*{.5cm}By including recoil effects (in the phase space factor, too), the rate of neutron beta decay can be
written as \cite{gk}:
\begin{eqnarray}
R^0 &=& \frac{G_0^2m_e^5}{2\pi^3} |V_{ud}|^2 \left[ 1.63387f_1^2  +
2.3775 \cdot 10^{-6} f_1f_2 + 1.1625\cdot 10^{-3}f_1f_3 +\right. 
\nonumber \\
&& \left. 1.5836 \cdot 10^{-6} f_2^2+ 3.1835
\cdot 10^{-6}f_1\lambda_{f_1} + 4.90159 g_1^2-6.2017\cdot 10^{-6}g_2^2 
 \right. 
\nonumber\\
&& \left. -3.1697 \cdot 10^{-7}g_1g_3-1.0161 \cdot 10^{-2}g_1g_2 
+ 1.1453 \cdot 10^{-5}g_1\lambda_{g_1} 
\right] \ , 
\label{deca} 
\end{eqnarray} 
where $m_e$ is the mass of the electron and $f_i,\ g_i$ are form factors
evaluated at zero momentum transfer. The coefficients $\lambda_{f_i},\
\lambda_{g_i}$ denote the corrections due to $q^2$ dependence of $f_i$ and 
$g_i$. We do not display the contributions of $f^2_3$, $g^2_3$, 
$\lambda_{f_2}$, $\lambda_{g_2}$ and $\lambda_{g_3}$ since their coefficients 
are at the $10^{-8}$ level, i.e., well below the current experimental
precision. In Eq.(\ref{deca}) we improved the precision of the coefficients of
the leading form factors by one more decimal place, with respect to Ref.
\cite{gk}.\\
\hspace*{.5cm}In order to get a simpler expression that incorporates the terms in
Eq. (\ref{deca}) relevant at order $10^{-4}$ let us discuss the size of the
different contributions. As is well known, in the limit of exact isospin
symmetry the conserved current hypothesis (CVC) allows one to obtain
 $f^{CVC}_1=1$ and $f^{CVC}_2=1.8529$ \cite{gk}. Also, exact G-parity, that is
expected to hold in the isospin symmetry  limit, predicts the
absence of second class form factors: $f_3=0$ and $g_2=0$. The
contribution of the axial form factor $g_3$ to the decay rate is expected to be below the $10^{-4}$
level, even if, according to the PCAC hypothesis, 
$g_3= (2m_n^2/m_{\pi}^2)g_1 \approx 114$.\\
\hspace*{.5cm}When isospin symmetry gets broken (due to the mass difference 
of $u$ and $d$ quarks), the above predictions change. For example, 
$f_1(0)$ will deviate slightly from unity while $g_2$ and $f_3$ become
different from zero. An estimate of these effects
indicates that $1-f_1 \approx 5 \times 10^{-5}$ \cite{paver}, $g_2
\approx 6 \times 10^{-3}$ and  $f_3 \sim 10^{-3}$ \cite{lee}. Therefore, the 
expression for the decay rate at the lowest order that is precise at 
the $10^{-4}$ level gets the simple form:
\begin{equation}
R^0 = \frac{G_0^2m_e^5}{2\pi^3} |V_{ud}|^2 (1.6335) (1+3\lambda^2)\ .
\label{ro}
\end{equation}

{ \large \bf 3. Radiative corrections.} An expression for the decay rate that 
is precise at this level requires the inclusion of radiative 
corrections up to order $\alpha^2$. The formal perturbative expansion in 
powers of $\alpha$ for the decay rate can be expressed as follows:
\begin{equation}
R=R^0 + R(\alpha) + R(\alpha^2) + \cdots
\label{alpha}
\end{equation}
where $R^0$ is the rate at tree-level discussed in the previous section
and $R(\alpha^n)$ denote its corrections of order $\alpha^n$.\\
\hspace*{.5cm}At any given order in $\alpha$, the radiative corrections can be
split into two terms: $R(\alpha^n) = R_{MI}(\alpha^n)+R_{MD}(\alpha^n)$, 
the model-independent (MI) and the model-dependent (MD) pieces, following 
their independence or not of the details of weak and strong interactions.\\
\hspace*{.5cm}The MI radiative corrections, also called {\it inner} corrections, arise
from long-distance QED (virtual and bremsstrahlung) corrections up to an
energy scale of $O({\rm 1\ GeV})$. The MD radiative corrections, also 
called {\it outer} corrections, involve essentially the  short-distance 
electroweak corrections.\\
\hspace*{.5cm}In the following sections we will discuss in detail the leading
orders of radiative corrections that have been computed in the
literature.\\
{\it \bf 3.1 Corrections of order $\alpha$.} The corrections of order $\alpha$ have 
been computed long ago \cite{sirlin67,sirlin78}. The model-independent piece of 
$O(\alpha)$ was given as \cite{sirlin67}
\begin{equation}
R_{MI}(\alpha)= \frac{G_F^2m_e^5}{2\pi^3} |V_{ud}|^2 (1+3\lambda^2)
\frac{1}{m_e^5} \int \left(f_{\alpha}+\frac{\alpha}{2\pi}g(E)\right) dS\ ,
\label{modin}
\end{equation}
while  the model-dependent contribution as \cite{sirlin78}
\begin{eqnarray}
R_{MD}(\alpha)&=& \frac{G_F^2m_e^5}{2\pi^3} |V_{ud}|^2 (1+3\lambda^2)   
\frac{1}{m_e^5} \int \frac{\alpha}{2\pi} \left( 3\ln \frac{m_W}{m_p}+
\ln \frac{m_W}{m_A}+\right.
\label{modde}\\
&&\left. 2C-4\ln \frac{m_W}{m_Z}+A_g \right) dS\ .
\nonumber
\end{eqnarray}
In the above equations $E$ denotes the energy of the electron,
$l=|\vec{l}|$ its
three-momentum and $dS\equiv El(E_m-E)^2dE$ . The function $f_{\alpha}$
denotes the coulombic correction
of order ${\alpha}$,
\begin{equation}
f_{\alpha} =\frac{\alpha}{\pi} \frac{\pi^2E}{l}\ ,
\label{fer}
\end{equation}
and the expression for $g(E)$ can be found in Ref. \cite{sirlin67}. 
The origin of the different terms in Eq. (\ref{modde}) can be found elsewhere
\cite{sirlin78,ms1}. Numerically, the MI radiative corrections at this 
order are
\begin{equation}
\frac{1}{m_e^5} \int f_{\alpha}dS = 0.0546\ ,~~~~~~~~~~~~~~~~~~~~~~~~~
~~~~~~~~~~~~~~~~~~~~~~~~~~~~~~~~~~~~~ 
\label{e10}
\end{equation} 
\begin{equation}
\frac{1}{m_e^5} \int \frac{\alpha}{2\pi} g(E) dS=0.0246 \ .~~~~~~~~~~~~~~~~~~~~~~
~~~~~~~~~~~~~~~~~~~~~~~~~~~~~~~~~
\label{e11}
\end{equation}
\hspace*{.5cm}As we can observe, the dominant piece of MI radiative corrections of order
$\alpha$ comes from the coulombic term. We will leave  the numerical
evaluation of the MD corrections to the following subsection.\\
\hspace*{.5cm}Before we close this subsection let us emphasize that the remaining
effects of radiative corrections that have a universal nature have
been absorbed in the redefinition of the effective weak coupling constant
with the result $G_0 \rightarrow G_F$, where $G_F=1.16639(2) \times
10^{-5}$ GeV$^{-2}$ can be extracted from muon decay \cite{pdg98}.\\
{\it \bf 3.2 Corrections of order $\alpha^2$.} In this subsection we consider 
the MI radiative corrections of $O(\alpha^2)$ \cite{sirlin87}. The MD 
corrections of this and higher orders will be
discussed in subsection 3.3. In a similar way to the previous subsection, we
can write the MI corrections as \cite{wilkinson1}:
\begin{equation}
R_{MI}(\alpha^2)= \frac{G_F^2m_e^5}{2\pi^3} |V_{ud}|^2 (1+3\lambda^2)   
\frac{1}{m_e^5} \int(f_{\alpha^2}+\delta_{\alpha^2} ) dS\ ,
\end{equation}
where again, we have separated explicitly the coulombic correction
$f_{\alpha^2}$ from the remaining corrections $\delta_{\alpha^2}$ at this
order.\\
\hspace*{.5cm}Numerically, we have:
\begin{eqnarray}
\frac{1}{m_e^5} \int f_{\alpha^2}dS = 0.0014\ , 
\label{e13} \\
\frac{1}{m_e^5} \int \delta_{\alpha^2} dS=0.0007 \ .
\label{e14}
\end{eqnarray}
As in the case of order $\alpha$,  we observe the dominance of
the coulombic piece in the MI radiative corrections of order $\alpha^2$.\\ 
{\it \bf 3.3 Corrections of order higher than $\alpha^2.$} The leading pieces 
of MI radiative corrections of order $\alpha^3$ have been computed in 
Ref. \cite{sirlin87}. Following the pattern already observed, we can expect 
that the MI corrections will be dominated also by the coulombic term. Indeed, 
coulombic corrections can be included to all orders (via the Fermi function
 $F(Z,E)$ \cite{Halpern}) by solving the
non-relativistic Schr\"odinger equation for the electron in the
Coulomb field of the proton. However, it it clear that for the purposes of our 
analysis, we have to keep only the MI corrections up to $O(\alpha^2)$.\\
\hspace*{.5cm}Now, we will focus on the MD radiative corrections.
Following the procedure discussed in \cite{sirlin78,ms1}, let us rewrite
the MD corrections of $O(\alpha)$  in the form
\begin{eqnarray}
R_{MD}(\alpha)&=&\frac{G_F^2m_e^5}{2\pi^3}|V_{ud}|^2 (1+3\lambda^2)
\frac{1}{m_e^5} \int \frac{\alpha}{2\pi} \left( 4\ln \frac{m_Z}{m_p}+
\ln \frac{m_p}{m_A}\right.
\nonumber\\
&&\left. \frac{}{}~~~~~~~~~~~~~~~~~~~~~~~~~~~~~~~~~~~~~~~~~
+2C+A_g \right) dS\ .
\label{sir}
\end{eqnarray}
The term $\ln(m_Z/m_p)$ is by far the dominant piece of this correction.
As in the case of coulombic corrections, one may wonder about the
relative size of higher order MD corrections having logarithmic terms. 
Ref. \cite{ms1} has pointed out that the effects of higher order can be
approximated by resummation of the leading-logarithmic corrections of
$O(\alpha^n \ln^2(m_Z/m_p)),\ n=1,2,\cdots$ via the renormalization group
equation (RGE). The net effect of this resummation is to replace Eq. (\ref{sir})
by \cite{ms1}:
\begin{eqnarray}
R'_{MD}&=& \frac{G_F^2m_e^5}{2\pi^3} |V_{ud}|^2 (1+3\lambda^2)
\frac{1}{m_e^5} \int\left[ \frac{\alpha}{2\pi} \left(\ln \frac{m_p}{m_A}+
2C+A_g \right) \right. \nonumber \\ 
&& \ \ \ \ \ \ \ \ \ \ \left. + S(m_p,m_Z)-S(m_p,m_p) \frac{}{} \right] dS\ ,
\end{eqnarray}
where $R'_{MD}$ denotes the MD radiative corrections that take into
account dominant logarithmic corrections at higher orders.\\ 
\hspace*{.5cm}According to ref. \cite{ms1}, $S(m_p,m_Z)$ can be obtained by solving an
specific RGE subject to the boundary condition condition $S(m_p,m_p)=1$.
The numerical result for this correction is:
\begin{equation}
S(m_p,m_Z)= 1.0220\ .
\label{e17}
\end{equation}
To obtain this numerical value, which differs slightly from Ref.
\cite{ms1}, we have used the fact that the top quark is heavier than the
$Z$ boson and should not be included in the thresholds when solving the
RGE.\\
{\it \bf 3.4 Final remarks on radiative corrections.} If we collect the 
results presented in the previous sections, we can write the radiative 
corrections to the neutron decay rate as
\begin{equation}
\sum_n R(\alpha^n)=\frac{G_F^2 m_e^5}{2\pi^3} |V_{ud}|^2(1+3\lambda^2)
 \left\{
\frac{1}{m_e^5} \int \left [  f_{\alpha} + \frac{\alpha(m_p)}{2\pi}g(E)
+f_{\alpha^2} \right. \right.~~  
\label{todo}
\end{equation}
\[
+\delta_{\alpha^2}+\frac{\alpha}{2\pi}\left[\ln\left(
\frac{m_p}{m_A}\right)+2C\right]+\frac{\alpha(m_p)}{2\pi}A_g\left.\left. 
+ S(m_p,m_Z)-S(m_pm_p) \frac{}{}\right] dS\right\}\ .
\nonumber
\]
\hspace*{.5cm}In some terms of the above equations we have replaced 
$\alpha^{-1}(0) =
137.036 \rightarrow \alpha^{-1}(m_p)=133.93$. According to Ref. \cite{ms1}  
$m_p$ is the relevant scale for the electromagnetic coupling in the 
renormalization scheme adopted in these calculations.\\
\hspace*{.5cm}Let us stress that  Eq. (\ref{todo}) has the {\it correct additive} structure in
terms of the expansion parameter $\alpha$. This observation is very
important  in view of several factorization
approximations currently found in the literature \cite{wilkinson1,towner}
which cannot be justified when the decay rate is computed at order  
$10^{-4}$. For example, it is customary to take the coulombic corrections to 
all orders through  the Fermi function as a global factor in the 
integrand of Eq. (\ref{todo}) \cite{wilkinson1}. This would lead to the 
following expression for the decay rate
\[
R=\frac{G_F^2 m_e^5}{2\pi^3} |V_{ud}|^2 (1+3\lambda^2)  \left\{
\frac{1}{m_e^5} \int F(Z,E)\left[ 1 + \frac{\alpha(m_p)}{2\pi}g(E)
+ \delta_{\alpha^2} \right. \right.
\] 
\begin{equation}
\left. \left. + \frac{\alpha}{2\pi} \left[ \ln \left(\frac{m_p}{m_A}
\right) +
2C \right] +\frac{\alpha(m_p)}{2\pi}A_g + S(m_p,m_Z)-S(m_p,m_p) \right] dS
\right\}
\end{equation}
\hspace*{.5cm}Clearly, this factorization introduces spurious terms of order 
$\alpha^2$, through the product of the term of order $\alpha$ in the Fermi 
function with other contributions of $O(\alpha)$ in the remaining corrections. 
A similar argument can be applied to the factorization of outer and inner 
radiative corrections (see ref. \cite{ms1} and Eq. (\ref{modde}) in 
Ref. \cite{towner}). Therefore, in the computation of the decay rate at the 
$10^{-4}$ level (or to a higher accuracy) radiative corrections must be taken 
in an additive form as shown in Eq.(\ref{alpha}) in 
order to avoid spurious corrections of order $\alpha^2$ (and higher orders).\\
\hspace*{.5cm}After these remarks, we proceed to obtain a numerical
expression for the decay rate involving radiative corrections.
The numerical values of some of these corrections have been provided in
Eqs. (\ref{e10},\ref{e11}, \ref{e13}, \ref{e14} and \ref{e17}). Currently, 
the largest uncertainty in radiative corrections comes from the terms which 
originate in the axial-induced photonic corrections,
\begin{equation}
\Delta \equiv  \frac{\alpha}{2\pi} \left[ \ln \left(\frac{m_p}{m_A}
\right) +
2C \right]\ .
\label{delta}
\end{equation}
\hspace*{.5cm}The non-asymptotic piece $C$ has been re-evaluated in 
ref. \cite{towner} with the result $C=0.881 \pm 0.030$. The mass parameter 
$m_A$ is taken as a low energy cutoff applied to the evaluation of the axial 
part of the asymptotic $\gamma W$ box correction. Marciano and Sirlin 
\cite{ms1} have suggested that, given the ambiguities in the choice of this 
cutoff, it is more appropriate to take the range $400\ {\rm MeV} \leq m_A \leq
1600\ {\rm MeV}$ for this axial mass. This wide range, is at present the
major source of uncertainty in radiative corrections to neutron beta
decay, since the term given in Eq. (\ref{delta}) becomes
\[
\Delta =0.0022\pm 0.0008 \ .
\]
\hspace*{.5cm}If we insert the different values of radiative corrections 
discussed above, we obtain an expression of the decay rate that is accurate at 
the $10^{-4}$ level:
\begin{equation}
R=0.1897 |V_{ud}|^2(1+3\lambda^2)(1+0.0739\pm 0.0008)\times 10^{-3} \
{\rm s}^{-1}\ ,
\label{e21}
\end{equation}
where the quoted error comes from the
uncertainty in the axial cutoff parameter \cite{ms1,towner}.\\
\hspace*{.5cm}Notice that if we had included radiative corrections with the 
Fermi function $F(Z,E)$ and MD corrections 
in a factorized form, we would have obtained,
\begin{equation}
R=0.1897 |V_{ud}|^2(1+3\lambda^2)(1+0.0758 \pm 0.0008) \times
10^{-3}\ {\rm s}^{-1}\ ,
\end{equation}
to be compared to Eq. (\ref{e21}).\\
\hspace*{.5cm}As it was previously mentioned, the factorization of some pieces 
of radiative corrections affects the expression for the decay rate at the 
order $10^{-4}$. Therefore, in the following we will use our Eq. (\ref{e21}) to quote a determination of $|V_{ud}|$.\\
{ \large \bf 4. Determination of $|V_{ud}|$. Comparison with other
sources.} According to Eq. (\ref{e21}), two experimental inputs are required to determine
$|V_{ud}|$ from neutron beta decay: the neutron lifetime $\tau_n\equiv R^{-1}$ 
and $\lambda$. We first summarize here 
the current situation regarding their measurements and afterwards we discuss
$|V_{ud}|$.\\
\hspace*{.5cm}{\it \bf 4.1 Neutron lifetime.} The world average reported by the 
Particle Data Group in the 1998 edition is \cite{pdg98}
\begin{equation}
\tau_n^{PDG98}=\frac{1}{R^{exp}}=(886.7 \pm 1.9)\ {\rm s}\ .
\label{taue}
\end{equation}
\hspace*{.5cm}This value is obtained \cite{pdg98} from a weighted average of ten
independent measurements and the quoted error has been inflated by a scale
factor S=1.2. A more recent and more precise measurement has been reported
in the literature \cite{th1,reich,arzu}: $\tau_n=(885.4\pm 1.0)\ {\rm s}$,
which is consistent with Eq. (\ref{taue}). Therefore, from its  weighted 
average with Eq. (\ref{taue}), we obtain
\begin{equation}
\tau_n^{new}=(885.9 \pm 0.9)\ {\rm s}\ .
\label{e23}
\end{equation}
\hspace*{.5cm}The set of measurements of $\tau_n$ is consistent as reflected by
a low value of the scale factor (S=1.07). As it was mentioned in the
introduction, the measurement of $\tau_n$ has reached the impressive
accuracy of 0.1 \%. \\
{\it \bf 4.2 Ratio $\lambda=g_1(0)/f_1(0)$.} The consistency of  measurements of 
$\lambda$ is not clear at present. So far,  measurements of the electron spin 
asymmetry  have provided the most precise values reported for $\lambda$. By 
combining the four most precise measurements (typically those with an accuracy 
higher than 0.4\%) one gets the weighted average \cite{pdg98}
\begin{equation}
\lambda^{PDG98}=1.2670 \pm 0.0035\ ,
\label{e25}
\end{equation}
where the error has been increased by an scale factor of $S=1.9$.
Therefore, the choice of this set of measurements would indicate a
precision of 0.3 \% in $\lambda$. Other less precise measurements may be
excluded because they have negligible statistical weight at the $10^{-4}$
level.\\  
\hspace*{.5cm}In order to study the impact of different measurements of
$\lambda$ on the determination of $|V_{ud}|$, we prefer to quote the
value of this parameter as a function of different values of $\lambda$.
Following Ref. \cite{pdg98}, we will choose the four most precise
measurements of $\lambda$: $\lambda_i$, where $i=L, B, Y, R$ denote,
respectively, the results of references \cite{liaud,bopp,yero,reich}.
We will exclude from our analysis the result of Ref. \cite{abele}, because
it has been superseded by the new result in Ref. \cite{reich}. The explicit
values are given in Table 1. The updated average becomes
\begin{equation}
\lambda^{PDG99}=1.2687 \pm 0.0016\ ,
\label{e26}
\end{equation} 
where the error has not been increased by any scale factor.\\
\hspace*{.5cm}Several comments about Eq. (\ref{e26}) are in order. First, the 
two most precise values of $\lambda$ used to obtain Eq. (\ref{e26}), namely, 
$\lambda_Y$ and $\lambda_R$ differ each one only by two standard deviations from 
Eq.(\ref{e26}) and by 4.7 standard deviations from each other. This reflects an 
inconsistency between both measurements. Second, the same inconsistency can be
observed in the ideogram shown in p. 622 of ref. \cite{pdg98}
where the set of 4 measurements used to quote Eq.(\ref{e25}) are  grouped into 
two separated peaks of Gaussian distributions. This signals the lack of
meaning of the single weighted average given in (\ref{e25}) and points out 
towards the  necessity of new and independent measurements of $\lambda$ to
elucidate the nature of the systematic errors.\\
{\it \bf 4.3 Determination of $|V_{ud}|$ from neutron beta decay.} In Table 1, 
we show the values of $|V_{ud}|$ obtained for individual measurements of 
$\lambda$ and some of its partial averages. Also displayed are the
corresponding values of the associated scale factor
$S=\sqrt{\sum_i\chi_i^2/(n-1)}$ when more than one value of $\lambda$ is
used; the bigger the scale factor is, the larger the inconsistency
within the set of chosen data. The inconsistency among the measurements of
$\lambda$ can be appreciated in several ways (see Table 1): $(a)$ the
central values of $|V_{ud}|$ corresponding to $\lambda_Y$ and $\lambda_R$
differ by almost 4$\sigma$, $(b)$ if we use the two sets of data
that appear clustered into two separated Gaussian distributions (see
section 4.2), namely, $\lambda_{LYB}$ and $\lambda_R$, we find that the 
corresponding central values of $|V_{ud}|$ differ by more than 4$\sigma$.\\
\hspace*{.5cm}We observe that in the latter case ($b$) the corresponding values of 
$|V_{ud}|$ violate the unitarity of the CKM matrix. Indeed, if we use 
$|V_{us}|=0.2196 \pm 0.0023$ and $|V_{ub}|= 0.0033 \pm 0.0008$ \cite{pdg98} 
we obtain (for comparison we include the unitarity check for SFT) 
\begin{equation}
|V_{ud}|^2+|V_{us}|^2+|V_{ub}|^2= \left \{
\begin{array}{l} 
0.9928\pm 0.0031, {\rm for}\ \lambda_R , \\
0.9969\pm 0.0022, {\rm for} SFT, \\
1.0066\pm 0.0035, {\rm for}\ \lambda_{LYB} \ .
\end{array} \right.
\label{unit}
\end{equation}
\hspace*{.5cm}It is very interesting to observe that when we combine the set of
four data, 
namely, if we use $\lambda_{LBYR}$, we obtain a value of $|V_{ud}|$
that is (accidentally!) consistent with unitarity 
$|V_{ud}|^2+|V_{us}|^2+|V_{ub}|^2=0.9986\pm 0.0025\ .$\\
\hspace*{.5cm}Therefore, since the whole set of most precise measurements of $\lambda$ is
inconsistent,  we can only quote two different values of $|V_{ud}|$, namely,
those corresponding to the two clearly separated Gaussian distributions
shown in the ideogram of p. 622 in Ref. \cite{pdg98}. Explicitly (for comparison
we include $|V_{ud}|$ from the unitarity of CKM matrix)
\begin{equation}
|V_{ud}| = \left\{ \begin{array}{l}
           0.9719 \pm 0.0019\pm0.0004\ {\rm from}\ \lambda_R , \\
           0.9790 \pm 0.0016\pm0.0004\ {\rm from}\ \lambda_{LYB}. \\
	   0.9756 \pm 0.0005\ {\rm from~unitary~of~CKM~matrix}. \\
          \end{array} \right.
\label{e29}	  
\end{equation}
\hspace*{.5cm}The first quoted error arises mainly from the uncertainty in the
measurement of $\lambda$, while the second one comes from the uncertainty in
the calculation of MD radiative corrections. This result indicates that
besides the interest of new measurements of $\lambda$ to solve the current 
discrepancy, they are also required to provide the most precise 
determination of $|V_{ud}|$.\\ 
\hspace*{.5cm}One may wonder where future measurements of $\lambda$ should lie 
if no new physics effects are relevant in neutron beta decay. In Fig. 1. we 
have plotted the one standard deviation elliptical contours of $|V_{ud}|$ and 
$\lambda$ obtained from Eq. (\ref{e21}) and the two above values of $\lambda$. 
For comparison we also display the two horizontal bands corresponding  to 
$|V_{ud}|$ from unitarity and SFT at one standard deviation, too.\\
\hspace*{.5cm}The small ellipse in Fig. 1 within the CKM band is the one
standard deviation region allowed by the unitarity $|V_{ud}|$ and 
Eq. (\ref{e23}) and it is the region where no new physics can yet appear in 
neutron beta decay. The current situation is that the two values of the $|V_{ud}|$ from neutron beta
decay and its value from SFT do not overlap with that region. Thus, Fig. 1.
provides a visualization of Eqs. (\ref{unit}) and (\ref{e29}). At any rate, neither of these discrepancies
with unitarity has enough statistical significance to be currently accepted as
an indication of the existence of new physics. But, the smallness of this 
region illustrates the potential of neutron beta decay to either
detect new physics or to provide strong lower bounds on its existence.\\
{ \large \bf 5. Conclusions.} The determination of $|V_{ud}|$ from neutron beta decay has 
several advantages. First, isospin breaking effects are well under control since
they affect the decay rate at the level below the $10^{-4}$ (contrary to
SFT where they enter at a few tenths of percent level \cite{towner}).
Second, the piece of MD radiative corrections $C$ (see Eq. (\ref{delta})) which is
associated to the non-asymptotic axial-induced photonic corrections only
receive contributions from one-body terms (contrary to SFT where the
photon and the $W$ gauge boson appearing in the box diagram can be coupled
to different nucleons). Third, possible residual effects of $Z$-dependent
radiative corrections (see Ref. \cite{wilkinson2}) are absent in neutron
beta decay. In addition, the non-leading form factors are irrelevant at this
level of precision. The complications of determining $g_1(0)$ do remain.
Fortunately, this form factor can be taken from experiment, because its
experimental value can be made appreciably more precise than its theoretical
one. Thus, the main source of theoretical uncertainties in the neutron beta
decay rate are the radiative corrections. We have seen that they are well under
control up to a few parts in $10^{-4}$. Prescriptions used to incorporate
different contributions of these corrections become relevant. By this we mean
that the (commonly used) factorization approximations of some pieces of
radiative corrections introduce spurious contributions at the $10^{-4}$ level.
The current prescription is the additive one.\\
\hspace*{.5cm}However, new measurements of $\lambda$ are required, first, to solve the
discrepancy among existing data and, second, to improve the determination of
$|V_{ud}|$. Actually (see Eq. (\ref{e29})), an improvement of a reduction of a
factor 2 in the error bars of $\lambda$ would provide a determination of $|V_{ud}|$
that is competitive with the one coming from SFT.\\
\hspace*{.5cm}The main conclusion in the above analysis is that the theoretical decay rate of
neutron beta decay is well under control up to a few parts at the $10^{-4}$ 
level. Given the present discrepancy of experimental measurements of $\lambda$
and our limitation to combine those data in a statistically significant
way, we can only quote the two favored values of $|V_{ud}|$ as given in
Eq. (\ref{e29}). These values correspond to the two subsets of consistent data
according to the ideogram shown in p. 622 of the Particle Data Group
\cite{pdg98}. New physics  effects can only be detected in neutron
beta decay outside the small ellipse within the CKM matrix band in Fig. 1.\\
\hspace*{.5cm}From Eq.(\ref{e29}) it becomes clear that, once the error bars on $\lambda$ are
appreciably  reduced, it will be very important to reduce the theoretical error
bars. Therefore it would be very useful to invest some effort in better controlling
the model dependent piece of the radiative corrections in neutron beta decay.\\
\\
{\bf Acknowledgments.}\\
The authors wish to express their gratitude to CONACyT (M\'exico). (J.L.G.L.)
wishes to acknowledge a leave of absence from Universidad de Guadalajara 
(M\'exico).

\begin{center}
\begin{table}
\begin{tabular}{|c|c|c|}
\hline
  $\lambda$&scale factor&$|V_{ud}|$\\
\hline
$\lambda_L$=1.2660(40)&--&$0.9766\pm0.0026$\\
\hline
$\lambda_Y$=1.2594(38)&--&$0.9809\pm0.0025$\\
\hline
$\lambda_B$=1.2620(50)&--&$0.9792\pm0.0032$\\
\hline
$\lambda_R$=1.2735(21)&--&$0.9719\pm0.0015$\\
\hline
$\lambda_{LB}$=1.2644(31)&0.6&$0.9777\pm0.0021$\\
\hline
$\lambda_{LY}$=1.2625(28)&1.2&$0.9789\pm0.0019$\\
\hline
$\lambda_{YB}$=1.2604(30)&0.4&$0.9802\pm0.0020$\\
\hline
$\lambda_{LR}$=1.2719(19)&1.7&$0.9729\pm0.0014$\\
\hline
$\lambda_{YR}$=1.2702(18)&3.2&$0.9740\pm0.0013$\\
\hline
$\lambda_{BR}$=1.2718(19)&2.1&$0.9730\pm0.0014$\\
\hline
$\lambda_{LYB}$=1.2624(24)&1.6&$0.9790\pm0.0017$\\
\hline
$\lambda_{LBR}$=1.2707(17)&1.8&$0.9737\pm0.0012$\\
\hline
$\lambda_{YBR}$=1.2692(17)&2.5&$0.9746\pm0.0012$\\
\hline
$\lambda_{YLR}$=1.2695(17)&2.4&$0.9744\pm0.0012$\\
\hline
$\lambda_{LYBR}$=1.2687(16)&2.1&$0.9749\pm0.0012$\\
\hline
\end{tabular}
\caption{Values of $|V_{ud}|$ obtained from different values of $\lambda$ 
and $\tau=(885.9\pm 0.9)\ s$, via Eq. \ref{e21}.}
\end{table}
\end{center}

\begin{figure}
\begin{center}
\label{neutron}
\centerline{\epsfig{file=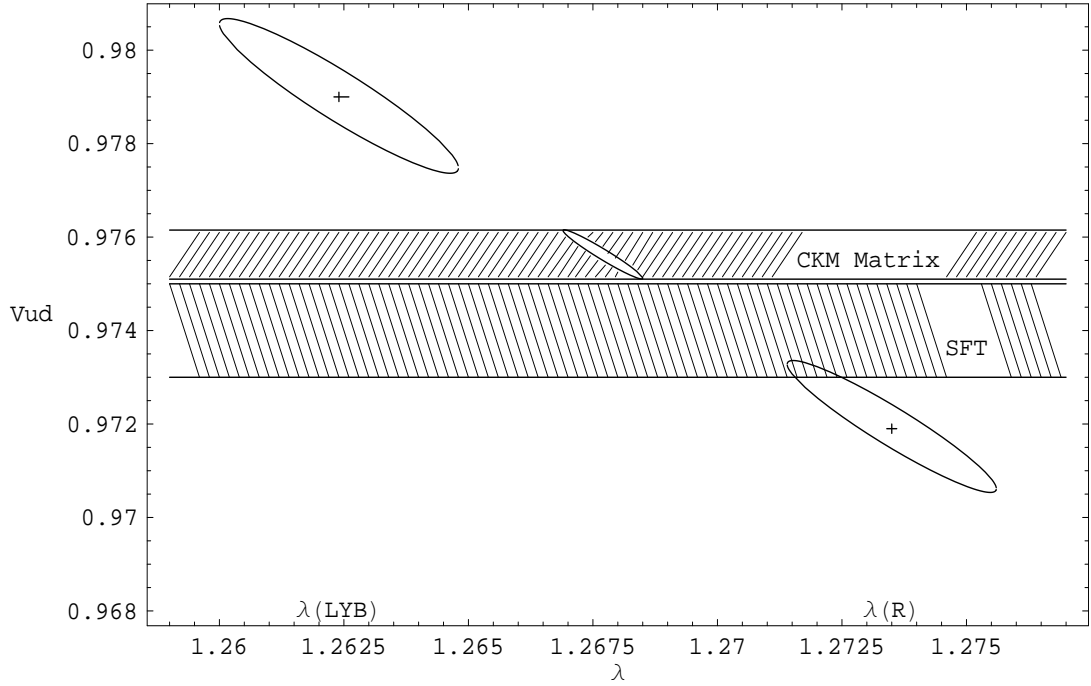,angle=0,width=16cm,height=10cm}}
\caption{$|V_{ud}|$ vs. $\lambda$. The two large ellipses corresponds to the 
$1\sigma$ contours obtained via Eq. (\ref{e21}). We also display the two
horizontal bands corresponding to $|V_{ud}|$ from unitarity of the CKM matrix 
and SFT at $1\sigma$, too. The small ellipse within the CKM matrix band illustrates the potential of
neutron beta decay to detect new physics or to put strong lower limits to its
existence.}
\end{center}
\end{figure}

\end{document}